\documentclass[letterpaper,spanish]{IEEEtran}

\usepackage[T1]{fontenc}
\usepackage[latin9]{inputenc}
\usepackage{color}
\usepackage{babel}
\addto\shorthandsspanish{\spanishdeactivate{~<>}}

\usepackage{array}
\usepackage{varioref}
\usepackage{graphicx}
\usepackage[unicode=true,
 bookmarks=true,bookmarksnumbered=false,bookmarksopen=false,
 breaklinks=true,pdfborder={0 0 1},backref=false,colorlinks=true]
 {hyperref}
\hypersetup{pdftitle={Implementación de Compilador Didáctico para pl0+},
 pdfauthor={Eduardo NAVAS},
 pdfsubject={Compiladores},
 pdfkeywords={compilación, fases de compilación}}

\makeatletter

\pdfpageheight\paperheight
\pdfpagewidth\paperwidth

\providecommand{\tabularnewline}{\\}

\newenvironment{lyxlist}[1]
	{\begin{list}{}
		{\settowidth{\labelwidth}{#1}
		 \setlength{\leftmargin}{\labelwidth}
		 \addtolength{\leftmargin}{\labelsep}
		 }}
	{\end{list}}

\makeatother

\usepackage{listings}
\begin{document}
\title{Implementación de un Compilador Didáctico para un súper-conjunto de
PL/0\thanks{Este artículo fue presentado en el \textbf{Congreso de Electrónca
e Informática 2010} de la \emph{Universidad Centroamericana ``José
Simeón Cañas''}, celebrado el 25 y 26 de noviembre de 2010.}}
\author{Eduardo Adam NAVAS-LÓPEZ, Catedrático del Departamento de Electrónica
e Informática, Universidad Centroamericana ``José Simeón Cañas'',
enavas@ing.uca.edu.sv}

\maketitle

\begin{abstract}
En este artículo se describen las características de un compilador
para un lenguaje súper-conjunto del conocido PL/0 creado por Niklaus
Wirth. La característica principal es que implementa las fases de
compilación de tal manera que la información que pasa entre cada una
se refleja como un archivo \textsf{XML}.
\end{abstract}


\section{Introducción}

\IEEEPARstart{A}{l} estudiar diseño de compiladores, siempre se habla
de las fases tradicionales de Análisis Léxico, Análisis Sintáctico,
Análisis Semántico y Generación de Código Objeto (ver \cite{Compiler Construction,Rochester,Programming Language Oberon,Project Oberon,Introducci=0000F3n a los compiladores,Libro del Drag=0000F3n}).
Conceptualmente estas fases son fáciles de diferenciar, sin embargo
los programadores normalmente no piensan en ellas. El comportamiento
por defecto de los compiladores tradicionales es ocultarlas para ofrecerle
a los programadores una respuesta rápida y efectiva. Esto es razonable
cuando lo que se quiere es un archivo ejecutable a partir de un conjunto
de archivos fuente.

Eventualmente, al estudiar sobre diseño y construcción de compiladores,
a los programadores les gustaría ver el proceso efectuado por las
fases de compilación de los compiladores que se usan tradicionalmente,
sin embargo, los compiladores tradicionales no permiten mostrar ese
tipo de información.

Así, este artículo presenta un compilador para un súper-conjunto de
PL/0, denominado aquí \textsf{pl0+}, que expone los datos que se transmiten
entre cada fase de compilación. También se demuestra que es posible
implementar un compilador en un lenguaje de alto nivel.\cite{c=0000F3digo}

\section{Metodología}

Las siguientes secciones describen brevemente el concepto, el diseño
y las fases del compilador denominado \texttt{tradukilo.py}.

\subsection{Diseño}

El compilador ha sido diseñado pensando no en su velocidad de ejecución,
sino en la posibilidad de usarlo para popósitos académicos-pedagógicos.

La idea es que el programador pueda elegir las fases de compilación
a ejecutar (algunas combinaciones por supuesto no son posibles). Por
ejemplo, si se desea estudiar la fase de análisis léxico, podría indicarse
al compilador que se detenga al terminar dicha fase. La salida del
compilador será entonces, un archivo en formato \textsf{xml} que contiene
la secuencia lineal de tokens que componen el código fuente.

Por otro lado, si ya se tiene un archivo \textsf{xml} que contiene
la especicación del árbol de sintaxis (que es la salida de la fase
de análisis sintáctico), podría indicársele al compilador que sólo
ejecute la fase de análisis semántico. La salida entonces será otro
archivo \textsf{xml} que contenga el árbol de sintaxis con las validaciones
de coincidencia de tipos y las otras tareas que realiza esta fase.

El funcionamiento por defecto del compilador deberá ser el de ejecutar
todas las fases de compilación, pero sin escribir en memoria secundaria
los flujos de comunicación de las diferentes fases.

\subsection{Lenguaje fuente}

El lenguaje fuente es un súper-conjunto del conocido lenguaje PL/0
creado por \emph{Niklaus Wirth}\cite{Alg+Data=00003DPrg}. Lo conoceremos
como lenguaje \textsf{pl0+} y su gramática se presenta a continuación:

\begin{lstlisting}[language=C,morekeywords={<programa>}]
<programa> ::= <bloque> '.'

<bloque> ::= 
  {<declaración_constantes>}? 
  {<declaración_variables>}? 
  {<declaración_procedimiento>}* 
  <instrucción> {';'}?

<declaración_constantes> ::= 'const'
  <identificador> '=' { '+' | '-' }? <número>
  { ','
    <identificador> '=' { '+' | '-' }? <número> 
  }* ';'

<declaración_variables> ::= 'var' <identificador>
  { ',' <identificador> }* ;

<declaración_procedimiento> ::= 'procedure'
  <identificador> ';' <bloque> ';'

<instrucción> ::= 
  <identificador> := <expresión> |
  'call' <identificador> |
  'begin' <instrucción> { ';' <instrucción> }*
    { ';' }? end |
  'if' <condición> 'then' <instrucción>
    { 'else' <instrucción> }? |
  'while' <condición> 'do' <instrucción> |
  'read' <identificador> |
  'write' <identificador> |
  <nada>

<condición> ::= 
  'odd' <expresión> |
  <expresión> { '=' | '<>' | '<' | '>' | '<=' 
    | '>=' } <expresión>

<expresión> ::= { '+' | '-' }? <término> 
  { { '+' | '-' } <término> }*

<término> ::= <factor> { { '*' | '/' } <factor> }*

<factor>::= { '-' }* { <identificador> | <número> | ( <expresión> ) }

<identificador> ::= <letra> { <letra> | <dígito> 
  | '_' }*
<número> ::= { <dígito> }+

<letra> ::= { 'a' | ... | 'z' }
<dígito> ::= { '0' | ... | '9' }

<Comentarios> ::= '(*'  <lo-que-sea>* '*)'
\end{lstlisting}

Es un lenguaje de programación sencillo de alto nivel que permite
anidamiento de procedimientos, recursión directa e indirecta, sólo
tiene variables y constantes del tipo de dato entero de 32\,bits
(es decir en el intervalo $\left[-2^{31},2^{31}-1\right]$) y tiene
los operadores aritméticos básicos y los relacionales para las condiciones.
Los procedimientos no retornan ningún valor, es decir que no hay funciones.
Y sólamente tiene instrucciones de entrada y salida básica de enteros
por la entrada estándar y la salida estándar respectivamente.

A continuación se presenta un programa de ejemplo que calcula los
números de la serie de Fibonacci:

\begin{lstlisting}[language=Pascal,caption={fibonacci.pl0+ -- Programa que muestra la serie de Fibonacci},label={fibonacci.pl0+}]
(*
    Cálculo de los números de la serie de fibonacci:
    f_0 = 1
    f_1 = 1
    f_n = f_{n-1} + f_{n-2}, n>1
*)
const fib_0=1, fib_1=1;
var n, f;

procedure fibonacci;
    var i,   (* Variable para hacer el recorrido *)
        f_1, (* Número Fibonacci anterior *)
        f_2; (* Número Fibonacci anterior al anterior *)
    begin
        if n=0 then f:=fib_0;
        if n=1 then begin
            f:=fib_1;
            write f; (* Los primeros dos elementos son iguales *)
        end;
        if n>1 then begin
            f_1:=1;
            write f_1;
            
            f_2:=1;
            write f_2;
            
            i:=2;
            while i<n do begin
                f:=f_1+f_2;
                write f;
                f_2:=f_1;
                f_1:=f;
                i:=i+1;
            end;
            f:=f_1+f_2;
        end;
    end;

begin
    read n;
    call fibonacci;
    write f;
end.
\end{lstlisting}

\subsection{Lenguaje objetivo}

El lenguaje objetivo es una variante del \textsf{código p} definido
para PL/0. Es un lenguaje tipo ensamblador y lo conoceremos en este
artículo como lenguaje \textsf{p+}. A continuación se presenta la
definición de sus instrucciones y mnemónicos:
\begin{lyxlist}{00.00.0000}
\item [{\texttt{LIT~<val>}}] Pone el literal numérico \texttt{<val>} en
el tope de la pila.
\item [{\texttt{CAR~<dif>~<pos>}}] ~\\
Copia el valor de la variable que está en la posición \texttt{<pos>}
en el bloque definido a \texttt{<dif>} niveles estáticos desde el
bloque actual en el tope de la pila.
\item [{\texttt{ALM~<dif>~<pos>}}] ~\\
Almacena lo que está en el tope de la pila en la variable que está
en la posición \texttt{<pos>} en el bloque definido a \texttt{<dif>}
niveles estáticos desde el bloque actual.
\item [{\texttt{LLA~<dif>~<dir>}}] \texttt{~}~\\
Llama a un procedimiento definido a \texttt{<dif>} niveles estáticos
desde el bloque actual, que comienza en la dirección \texttt{<dir>}.
\item [{\texttt{INS~<num>}}] Instancia un procedimiento, reservando espacio
para las \texttt{<num>} variables del bloque que lo implementa (este
número incluye las celdas necesarias para la ejecución del código,
que en el caso del lenguaje \textsf{p+}\footnote{igual que en el caso del \textsf{código p}}
son 3 enteros adicionales).
\item [{\texttt{SAC~<dir>}}] Salto condicional hacia la dirección \texttt{<dir>}
si el valor en el tope de la pila es 0.
\item [{\texttt{SAL~<dir>}}] Salto incondicional hacia la dirección \texttt{<dir>}.
\item [{\texttt{OPR~<opr>}}] Operación aritmética o relacional, según
el número \texttt{<opr>}. Los parámetros son los valores que están
actualmente en el tope de la pila y allí mismo se coloca el resultado.
Los valores posibles para \texttt{<opr>} son:

\begin{lyxlist}{00.00.0000}
\item [{1:}] Negativo (menos unario)
\item [{2:}] Suma (\texttt{+})
\item [{3:}] Resta (\texttt{-})
\item [{4:}] Multiplicación (\texttt{{*}})
\item [{5:}] División (\texttt{/})
\item [{6:}] Operador \emph{odd} (impar)
\item [{8:}] Comparación de igualdad (\texttt{=})
\item [{9:}] Diferente de (\texttt{<>})
\item [{10:}] Menor que (\texttt{<})
\item [{11:}] Mayor o igual que (\texttt{>=})
\item [{12:}] Mayor que (\texttt{>})
\item [{13:}] Menor o igual que (\texttt{<=})
\end{lyxlist}
\item [{\texttt{RET}}] Retornar de procedimiento.
\item [{\texttt{LEE}}] Lee un valor de la entrada estándar y la almacena
en el tope de la pila.
\item [{\texttt{ESC}}] Escribe en la salida estándar el valor del tope
de la pila.
\end{lyxlist}

\subsection{Interface del compilador}

La interface del compilador es por línea de comandos. La sintaxis
general para invocarlo es la siguiente:\\
\texttt{}
\begin{lstlisting}[frame=TBlr,frameround=ffff]
$ python tradukilo.py [-a] [-m] [-e] [--lex] [--sin] [--sem] [--gen] programa
\end{lstlisting}

Las opciones y sus significados son los siguientes:
\begin{lyxlist}{00.00.0000}
\item [{\texttt{-a}}] (\texttt{-{}-ayuda}) Muestra un mensaje de ayuda
y termina de inmediato.
\item [{\texttt{-m}}] (\texttt{-{}-mostrar}) En caso de haber una compilación
sin errores, muestra el resultado del proceso en la pantalla.
\item [{\texttt{-e}}] (\texttt{-{}-errores-xml}) Imprime los errores y
las advertencias en la salida estándar de error en formato de archivo
\textsf{xml}. Esta opción está destinada a reportar los errores y
advertencias de manera estructurada para ser procesados por una eventual
interfaz gráfica para el compilador.
\item [{\texttt{-{}-lex}}] Indica que se debe ejecutar la fase de análisis
léxico.
\item [{\texttt{-{}-sin}}] Indica que se debe ejecutar la fase de análisis
sintáctico.
\item [{\texttt{-{}-sem}}] Indica que se debe ejecutar la fase de análisis
semántico.
\item [{\texttt{-{}-gen}}] Indica que se debe ejecutar la fase de generación
de código objeto.
\end{lyxlist}
Al ejecutar el comando, se intenta compilar el archivo \texttt{programa}.
Si no se indica ninguna fase de compilación particular, se asumen
todas. Si la compilación tiene éxito, se genera un archivo con extensión
diferente, dependiendo de la última fase ejecutada.

La extensión de los programas \textsf{pl0+} se asume como \texttt{.pl0+},
la extensión de salida del análisis léxico se asume \texttt{.pl0+lex},
la extensión de salida del análisis sintáctico se asume \texttt{.pl0+sin},
la de análisis semántico \texttt{.pl0+sem} y la de la generación de
código objeto \texttt{.p+}.

Pueden ejecutarse combinaciones como:\\
\begin{lstlisting}[breaklines=true,frame=TBlr,frameround=ffff]
$ python tradukilo.py programa.pl0+
$ python tradukilo.py --lex programa.pl0+
$ python tradukilo.py --lex --sin programa.pl0+
$ python tradukilo.py programa.pl0+ --lex --sin --sem
$ python tradukilo.py programa.pl0+lex --sin    --sem
$ python tradukilo.py programa.pl0+sin --sem    --gen
\end{lstlisting}

De las líneas anteriores, la primera ejecuta todas las fases de compilación
sobre el archivo fuente \texttt{programa.pl0+} y genera un archivo
llamado \texttt{programa.p+}. La segunda sólo ejecuta la fase de análisis
léxico y genera un archivo llamado \texttt{programa.pl0+lex}. La tercera
ejecuta las fases de análisis léxico y sintáctico y genera un archivo
llamado \texttt{programa.pl0+sin}. La cuarta ejecuta las fases de
análisis léxico, sintáctico y semántico y genera un archivo llamado
\texttt{programa.pl0+sem}. La quinta toma un archivo \texttt{programa.pl0+lex}
con la lista de \emph{tokens} de un programa fuente, ejecuta las fases
de análisis sintáctico y semántico y genera un archivo llamado \texttt{programa.pl0+sem}.
La sexta toma un archivo \texttt{programa.pl0+sin} que contiene el
árbol de sintaxis de un programa fuente, ejecuta las fases de análisis
semántico y de generación de código objeto y genera un archivo llamado
\texttt{programa.p+}.

Evidentemente no todas las combinaciones son posibles. Por ejemplo,
no se puede solicitar ejecutar las fases de análisis léxico (\texttt{-{}-lex})
y de análisis semántico (\texttt{-{}-sem}). En tales casos, el compilador
responderá con un mensaje de error al usuario.

\begin{figure*}
\begin{centering}
\includegraphics[width=0.75\textwidth]{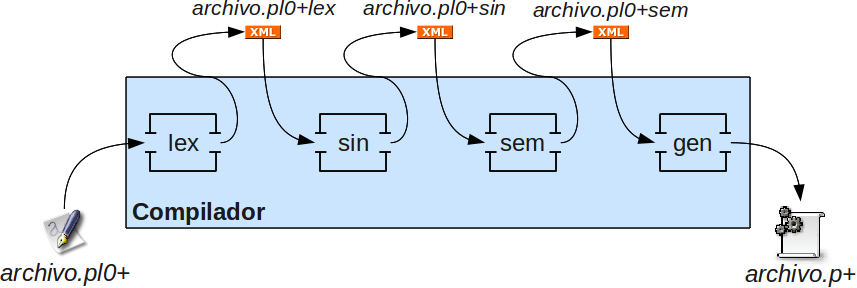}
\par\end{centering}
\caption{Esquema del compilador\label{fig:Esquema-del-compilador}}

\rule[0.5ex]{1\linewidth}{1pt}
\end{figure*}

Las extensiones de entrada y salida de cada fase se describen en la
siguiente tabla y en la figura \vref{fig:Esquema-del-compilador}:

\begin{tabular}{|>{\centering}p{1.5cm}|>{\centering}p{1.5cm}|>{\centering}p{2cm}|>{\centering}p{2cm}|}
\hline 
Código de la fase & Opción de la fase & Extensión entrada & Extensión salida\tabularnewline
\hline 
\hline 
\textbf{lex} & \texttt{-{}-lex} & \texttt{.pl0+} & \texttt{.pl0+lex}\tabularnewline
\hline 
\textbf{sin} & \texttt{-{}-sin} & \texttt{.pl0+lex} & \texttt{.pl0+sin}\tabularnewline
\hline 
\textbf{sem} & \texttt{-{}-sem} & \texttt{.pl0+sin} & \texttt{.pl0+sem}\tabularnewline
\hline 
\textbf{gen} & \texttt{-{}-gen} & \texttt{.pl0+sem} & \texttt{.p+}\tabularnewline
\hline 
\end{tabular}

Cabe recalcar que sólo se crea el archivo de la última fase de compilación
ejecutada, y no los intermedios.

\subsection{Adición de fases complementarias\label{subsec:Adici=0000F3n-de-fases}}

El compilador está implementado como un conjunto de programas Python~2.6
organizados de la siguiente manera:

\begin{lstlisting}[frame=none]
tradukilo.py  --> Interface principal del compilador

fases/
  __init__.py --> Configuración global
  lex.py      --> Análisis léxico
  sin.py      --> Análisis sintáctico
  sem.py      --> Análisis semántico
  gen.py      --> Generación de código objeto

rulilo.py     --> Intérprete de programas p+
\end{lstlisting}

En el archivo \texttt{\_\_init\_\_.py} contiene una lista llamada
\texttt{fasesDisponibles} que determina cuáles son las fases que pueden
invocarse desde la interfaz principal del compilador. Puede agregarse
una fase nueva, incluyendo las opciones de línea de comandos simplemente
agregando la información del nombre del módulo, el archivo en el que
está, el nombre de la función de traducción en ese módulo, la extensión
de salida de esa fase y una descripción para usarse en los mensajes
de error.

El contenido actual de la lista está incluído en el siguiente fragmento
de código:

\begin{lstlisting}[language=Python]
fasesDisponibles = [ {
    'nombre': 'lex',
    'archivo': 'lex.py',
    'función': 'faseTraduccion',
    'extensiónSalida': '.pl0+lex',
    'descripción': 'Fase de análisis léxico',
  }, {
    'nombre': 'sin',
    'archivo': 'sin.py',
    'función': 'faseTraduccion',
    'extensiónSalida': '.pl0+sin',
    'descripción': 'Fase de análisis sintáctico',
  }, {
    'nombre': 'sem',
    'archivo': 'sem.py',
    'función': 'faseTraduccion',
    'extensiónSalida': '.pl0+sem',
    'descripción': 'Fase de análisis semántico',
  }, {
    'nombre': 'gen',
    'archivo': 'gen.py',
    'función': 'faseTraduccion',
    'extensiónSalida': '.p+',
    'descripción': 'Fase de generación de código objeto',
  } ]
\end{lstlisting}

\subsection{Manejo y reporte de Errores}

En este compilador los errores se registran en dos listas internas
que se van pasando de fase en fase, diferenciando entre los ``\emph{errores}''
y las ``\emph{advertencias}''.

Los \emph{errores} son aquellos fragmentos de código que hacen imposible
la compilación porque es muy difícil o imposible determinar la intención
del programador.

Por otro lado, las \emph{advertencias} son fragmentos de código que
permiten la compilación, porque puede suponerse la intención del programador,
pero el programa fuente no es totalmente válido y el resultado de
la compilación puede no coincidir del todo con la verdadera intención
del programador.

Este compilador, así como muchos otros, muestra primero los \emph{errores}
y luego las \emph{advertencias} y en cada grupo, los errores están
ordenados respecto de su aparición en el código fuente.

A continuación se presenta un programa \textsf{pl0+} con múltiples
errores (programa \ref{errores.pl0+}) y luego la salida estándar
del compilador al intentar compilarlo (código \ref{errores.err}):

\begin{lstlisting}[language=Pascal,caption={errores.pl0+ -- Código con errores},label={errores.pl0+}]
const n=5;
var f, i;
begin
    f := 2  5 * 9
    i := 2 % f;
    f := 12*2+5*9/8;
    f := 8 / i * 2 -;
    f := 9 - i * 2
    if n<>1 then begin
        i:=2;
        while i 5 <= n 2 do begin
            f1:=f; i:=i+1;
        end
    end;
end.
\end{lstlisting}

\begin{lstlisting}[caption={Salida estándar al intentar compilar el programa anterior},label={errores.err}]
ERROR****************
Fase de origen:sin
Línea 4: Falta un operador
    f := 2  5 * 9
----------^
ERROR****************
Fase de origen:sin
Línea 5: Falta un operador
    i := 2 % f;
----------^
ERROR****************
Fase de origen:lex
Línea 5: Caracter inválido.
    i := 2 % f;
-----------^
ERROR****************
Fase de origen:sin
Línea 7: Se esperaba 'identificador', 'numero', '-' o '(...)', pero se encontró ';'
    f := 8 / i * 2 -;
--------------------^
ERROR****************
Fase de origen:sin
Línea 11: Falta un operador
        while i 5 <= n 2 do begin
---------------^
ERROR****************
Fase de origen:sem
Línea 12: Referencia a variable no declarada
            f1:=f; i:=i+1;
------------^
ADVERTENCIA**********
Fase de origen:sin
Línea 8: Falta un ';'
    f := 9 - i * 2
------------------^
\end{lstlisting}

Para permitir que el compilador pueda ser acoplado con un posible
editor especial de código fuente \textsf{pl0+}, que muestre los errores
léxicos, sintácticos y semánticos de los programas, los \emph{errores}
y \emph{advertencias} también se pueden desplegar en forma de archivo
\textsf{xml} estructurado a través de la salida estándar de error.

El siguiente archivo representa el mismo reporte de errores para el
programa \ref{errores.pl0+} pero habiendo incluído la opción \texttt{-e}
(o \texttt{-{}-errores-xml}) al intentar compilarlo:

\begin{lstlisting}[caption={Errores del programa errores.pl0+ en formato xml},label={errores.xml}]
<?xml version="1.0" ?>
<errores_y_advertencias>
  <errores>
    <error columna="10" linea="4">
      <mensaje>
        Falta un operador
      </mensaje>
      <contexto>
<![CDATA[    f := 2  5 * 9]]>      </contexto>
      <fase nombre="sin">
        Fase de análisis sintáctico
      </fase>
    </error>
    <error columna="10" linea="5">
      <mensaje>
        Falta un operador
      </mensaje>
      <contexto>
<![CDATA[    i := 2 % f;]]>      </contexto>
      <fase nombre="sin">
        Fase de análisis sintáctico
      </fase>
    </error>
    <error columna="11" linea="5">
      <mensaje>
        Caracter inválido.
      </mensaje>
      <contexto>
<![CDATA[    i := 2 % f;]]>      </contexto>
      <fase nombre="lex">
        Fase de análisis léxico
      </fase>
    </error>
    <error columna="20" linea="7">
      <mensaje>
        Se esperaba 'identificador', 'numero', '-' o '(...)', pero se encontró ';'
      </mensaje>
      <contexto>
<![CDATA[    f := 8 / i * 2 -;]]>      </contexto>
      <fase nombre="sin">
        Fase de análisis sintáctico
      </fase>
    </error>
    <error columna="15" linea="11">
      <mensaje>
        Falta un operador
      </mensaje>
      <contexto>
<![CDATA[        while i 5 <= n 2 do begin]]>      </contexto>
      <fase nombre="sin">
        Fase de análisis sintáctico
      </fase>
    </error>
    <error columna="12" linea="12">
      <mensaje>
        Referencia a variable no declarada
      </mensaje>
      <contexto>
<![CDATA[            f1:=f; i:=i+1;]]>      </contexto>
      <fase nombre="sem">
        Fase de análisis semántico
      </fase>
    </error>
  </errores>
  <advertencias>
    <error columna="18" linea="8">
      <mensaje>
        Falta un ';'
      </mensaje>
      <contexto>
<![CDATA[    f := 9 - i * 2]]>      </contexto>
      <fase nombre="sin">
        Fase de análisis sintáctico
      </fase>
    </error>
  </advertencias>
</errores_y_advertencias>
\end{lstlisting}

Finalmente, la estrategia general de procesamiento de errores consiste
en que cuando se encuentra un error, el compilador hace una serie
de suposiciones y ``salta'' hasta algún otro punto del programa
para intentar continuar con su trabajo.

En algunas ocaciones, esos conjuntos de suposiciones y saltos, no
son correctos, y el compilador se ``desestabiliza'', provocando
detección de errores falsos.

Para disminuir la frecuencia de las detecciones falsas de errores,
se optó por aplicar ---al menos en la fase de análisis sintáctico
al construir el árbol de operaciones de las expresiones--- la regla
eurística que indica que lo normal es que sólo haya un error en una
misma línea de código.

Esto, en términos de implementación se solucionó con que al momento
de ``identificar'' un error o advertencia, se consulta si ya hay
identificado un error o advertencia en la misma línea y de la misma
fase. Si lo hay, no se registra el más reciente.

\subsection{Interface del intérprete}

La interface del intérprete es por línea de comandos. La sintaxis
general para invocarlo es la siguiente:\\
\texttt{}
\begin{lstlisting}[frame=TBlr,frameround=ffff]
$ python rulilo.py [-a] [-d] programa-objeto.p+
\end{lstlisting}

Las opciones y sus significados son los siguientes:
\begin{lyxlist}{00.00.0000}
\item [{\texttt{-a}}] (\texttt{-{}-ayuda}) Muestra un mensaje de ayuda
y termina de inmediato.
\item [{\texttt{-d}}] (\texttt{-{}-depurar}) Ejecuta cada instrucción haciendo
una espera, para que el usuario observe el valor de los registros
de la máquina virtual.
\end{lyxlist}

\section{Conclusiones}
\begin{enumerate}
\item Disponer de un compilador que permita examinar el resultado de cada
fase de compilación, ayuda a comprender el propósito de estas.
\item Al implementar un compilador, la separación de las fases de compilación
---es decir, implementándolas como módulos funcionales separados
que se transfieren información entre sí de manera secuencial--- contribuye
a su mejor comprensión, permite una depuración más fácil y permite
ampliaciones futuras. Sin embargo, también exige documentar detalladamente
las transformaciones que sufre el programa entre ellas.
\item El diseño modular del compilador \texttt{tradukilo.py} permite que
este sea ampliado adicionando fases de optimización y procesamiento
adicional antes y después de las fases principales (ver sección \ref{subsec:Adici=0000F3n-de-fases}).
\item El formato \textsf{xml} permite el compartimiento genérico de información
entre diferentes componentes de software de manera simple y legible
por humanos.
\item Un compilador puede ser implementado en un lenguaje de alto nivel
(como Python) y no necesariamente implica que el compilador será lento
en la escala de tiempo humano.
\end{enumerate}

\section{Discución de las conclusiones y resultados}

Debido a la simplicidad y limitaciones del lenguaje \textsf{pl0+}
---no hay funciones, no hay parámetros en los subprocedimientos,
sólo hay un tipo de dato, no hay estructuras, no hay clases, las declaraciones
sólo pueden ir al principio de los bloques, etc.--- las tareas de
la fase de análisis semántico de \texttt{tradukilo.py} son muy limitadas
y escuetas, y sería más interesante ver el análisis semántico necesario
para lenguajes más ricos.

Hubo muchos aspectos sobre el tratamiento de errores que quedó fuera
de este artículo y que merecen ---por rigurosidad académica--- ser
presentados y analizados. Esto podría presentarse posteriormente en
otro artículo.

Luego de la implementación del intérprete de código \texttt{p+}, se
hace pantente que convendría hacer algunas reflexiones sobre las características
del lenguaje objetivo y sobre la implementación del intérprete correspondiente.
Por ejemplo pueden analizarse el lenguaje intermedio PIR y el de bajo
nivel PASM del proyecto \emph{Parrot} (ver \cite{Parrot}).

Finalmente, luego de haber implementado un compilador para un lenguaje
sencillo de alto nivel (\textsf{pl0+}), y un intérprete para un lenguaje
de bajo nivel (\textsf{p+}), surge la idea general de diseñar y construir
un conjunto de herramientas de propósito pedagógico, herramientas
propias (conocidas desde dentro) que sirvan para varios propósitos
en varios ámbitos dentro del quehacer académico de las ciencias de
la computación.

Por ejemplo, se puede diseñar un lenguaje de programación de alto
nivel orientado a los algoritmos en pseudocódigo\footnote{El Departamento de Electrónica e Informática ya ha iniciado un esfuerzo
similar y ya se tiene experiencia en esta área.}; diseñar un lenguaje de programación de bajo nivel, genérico, orientado
a código de máquina, en el que se pueda programar directamente (a
diferencia del lenguaje \textsf{p+}); diseñar y construir un compilador
que traduzca programas escritos del primer lenguaje al segundo y que
permita, así como \texttt{tradukilo.py}, examinar el resultado de
todas sus fases; y obviamente construir un intérprete de programas
en el segundo lenguaje.

Un conjunto tal de herramientas tendría múltiples propósitos u objetivos,
como los siguientes:
\begin{itemize}
\item Introducir los tópicos básicos de la algoritmia y la programación
de computadoras en alto nivel, a estudiantes de educación media, estudiantes
universitarios, y catedráticos que requieran de un formato genérico
y simple para describir algoritmos.
\item Introducir los conceptos básicos de programación de computadoras en
bajo nivel, a estudiantes de educación media en el área de electrónica
y estudiantes universitarios en el área de arquitectura de computadoras,
en un entorno de ejecución seguro, sin posibilidades de dañar los
equipos físicos y sin consecuencias colaterales por investigar o comenter
errores.
\item Contar con un compilador propio, hecho en casa, en el tercer mundo,
diseñado para ser leído y estudiado, y poder profundizar en el diseño
y construcción de ese tipo de herramientas o de piezas de software
similares.
\item Cumplir el importante propósito de contar con un amplio conjunto de
herramientas que puedan ser estudiadas desde diferentes perspectivas
a lo largo de una carrera de grado ---y tal vez también de postgrado---,
como la Licenciatura en Ciencias de la Computación que ofrece la Universidad
Centroamericana ``José Simeón Cañas''.
\end{itemize}
Yendo más allá ---y tal vez tan sólo pensando en voz alta---, podría
extrapolarse el desarrollo de estas herramientas a un sistema operativo
completo con propósitos académicos, de investigación y desarrollo,
como los presentados en \cite{Project Oberon} y \cite{Amoeba} que
tienen sus propios lenguajes de programación.

\section{Anexos}

\subsection{Descripción de la Fase de Análisis Léxico}

En cualquier compilador el propósito de esta fase es convertir los
caracteres del archivo que contiene al programa fuente, en una secuencia
lineal de los elementos mínimos con un significado en el lenguaje
(y sus eventuales valores). Estos elementos mínimos con significado
son llamados \emph{Elementos Léxicos} o \emph{Tokens}.\cite{Compiler Construction,Introducci=0000F3n a los compiladores,Libro del Drag=0000F3n}

Para cada elemento léxico de los programas en lenguaje \textsf{pl0+},
la fase de análisis léxico genera una etiqueta \textsf{xml} que lo
representa. Cada etiqueta tiene los atributos \texttt{columna}, \texttt{linea}
y \texttt{longitud}, que indican respectivamente la columna donde
comienza el elemento, la línea en que se encuentra y la longitud del
elemento léxico que representa.

Las palabras reservadas del lenguaje \textsf{pl0+} son: '\texttt{begin}',
'\texttt{call}', '\texttt{const}', '\texttt{do}', '\texttt{end}',
'\texttt{if}', '\texttt{odd}', '\texttt{procedure}', '\texttt{then}',
'\texttt{var}', '\texttt{while}', '\texttt{else}', '\texttt{write}'
y '\texttt{read}'. Cada una de ellas se representa con una etiqueta
con el mismo nombre pero en mayúsculas.

Los identificadores se representan con una etiqueta con nombre \texttt{IDENTIFICADOR}
con el atributo \texttt{nombre} que indica el nombre del identificador.

Los números que aparecen en los programas, se representan como etiquetas
con nombre \texttt{NUMERO} con el atributo \texttt{valor} que contiene
el valor del número.

Los demás elementos léxicos de \textsf{pl0+} se representan según
la siguiente tabla:

\begin{center}
\begin{tabular}{|c|c|}
\hline 
{\footnotesize{}Símbolo} & {\footnotesize{}Etiqueta}\tabularnewline
\hline 
\hline 
\texttt{\footnotesize{}=} & \texttt{\footnotesize{}igual}\tabularnewline
\hline 
\texttt{\footnotesize{}:=} & \texttt{\footnotesize{}asignacion}\tabularnewline
\hline 
\texttt{\footnotesize{},} & \texttt{\footnotesize{}coma}\tabularnewline
\hline 
\texttt{\footnotesize{};} & \texttt{\footnotesize{}punto\_y\_coma}\tabularnewline
\hline 
\texttt{\footnotesize{}(} & \texttt{\footnotesize{}parentesis\_apertura}\tabularnewline
\hline 
\texttt{\footnotesize{})} & \texttt{\footnotesize{}parentesis\_cierre}\tabularnewline
\hline 
\texttt{\footnotesize{}<>} & \texttt{\footnotesize{}diferente}\tabularnewline
\hline 
\texttt{\footnotesize{}<} & \texttt{\footnotesize{}menor\_que}\tabularnewline
\hline 
\texttt{\footnotesize{}>} & \texttt{\footnotesize{}mayor\_que}\tabularnewline
\hline 
\texttt{\footnotesize{}<=} & \texttt{\footnotesize{}menor\_igual}\tabularnewline
\hline 
\texttt{\footnotesize{}>=} & \texttt{\footnotesize{}mayor\_igual}\tabularnewline
\hline 
\texttt{\footnotesize{}+} & \texttt{\footnotesize{}mas}\tabularnewline
\hline 
\texttt{\footnotesize{}-} & \texttt{\footnotesize{}menos}\tabularnewline
\hline 
\texttt{\footnotesize{}{*}} & \texttt{\footnotesize{}por}\tabularnewline
\hline 
\texttt{\footnotesize{}/} & \texttt{\footnotesize{}entre}\tabularnewline
\hline 
\texttt{\footnotesize{}.} & \texttt{\footnotesize{}punto}\tabularnewline
\hline 
\end{tabular}
\par\end{center}

A continuación se presenta el resultado de aplicar el análisis léxico
sobre el programa \vref{fibonacci.pl0+}:

\begin{lstlisting}[caption={fibonacci.pl0+lex -- Elementos léxicos de fibonacci.pl0+},label={fibonacci.pl0+lex}]
<?xml version="1.0" ?>
<lexemas>
  <!--
**************************************
Por favor, no modifique este archivo
a menos que sepa lo que está haciendo
**************************************
-->
  <CONST columna="0" linea="7" longitud="5"/>
  <IDENTIFICADOR columna="6" linea="7" longitud="5" nombre="fib_0"/>
  <igual columna="11" linea="7" longitud="1"/>
  <NUMERO columna="12" linea="7" longitud="1" valor="1"/>
  <coma columna="13" linea="7" longitud="1"/>
  <IDENTIFICADOR columna="15" linea="7" longitud="5" nombre="fib_1"/>
  <igual columna="20" linea="7" longitud="1"/>
  <NUMERO columna="21" linea="7" longitud="1" valor="1"/>
  <punto_y_coma columna="22" linea="7" longitud="1"/>
  <VAR columna="0" linea="8" longitud="3"/>
  <IDENTIFICADOR columna="4" linea="8" longitud="1" nombre="n"/>
  <coma columna="5" linea="8" longitud="1"/>
  <IDENTIFICADOR columna="7" linea="8" longitud="1" nombre="f"/>
  <punto_y_coma columna="8" linea="8" longitud="1"/>
  <PROCEDURE columna="0" linea="10" longitud="9"/>
  <IDENTIFICADOR columna="10" linea="10" longitud="9" nombre="fibonacci"/>
  <punto_y_coma columna="19" linea="10" longitud="1"/>
  <VAR columna="4" linea="11" longitud="3"/>
  <IDENTIFICADOR columna="8" linea="11" longitud="1" nombre="i"/>
  <coma columna="9" linea="11" longitud="1"/>
  <IDENTIFICADOR columna="8" linea="12" longitud="3" nombre="f_1"/>
  <coma columna="11" linea="12" longitud="1"/>
  <IDENTIFICADOR columna="8" linea="13" longitud="3" nombre="f_2"/>
  <punto_y_coma columna="11" linea="13" longitud="1"/>
  <BEGIN columna="4" linea="14" longitud="5"/>
  <IF columna="8" linea="15" longitud="2"/>
  <IDENTIFICADOR columna="11" linea="15" longitud="1" nombre="n"/>
  <igual columna="12" linea="15" longitud="1"/>
  <NUMERO columna="13" linea="15" longitud="1" valor="0"/>
  <THEN columna="15" linea="15" longitud="4"/>
  <IDENTIFICADOR columna="20" linea="15" longitud="1" nombre="f"/>
  <asignacion columna="21" linea="15" longitud="2"/>
  <IDENTIFICADOR columna="23" linea="15" longitud="5" nombre="fib_0"/>
  <punto_y_coma columna="28" linea="15" longitud="1"/>
  <IF columna="8" linea="16" longitud="2"/>
  <IDENTIFICADOR columna="11" linea="16" longitud="1" nombre="n"/>
  <igual columna="12" linea="16" longitud="1"/>
  <NUMERO columna="13" linea="16" longitud="1" valor="1"/>
  <THEN columna="15" linea="16" longitud="4"/>
  <BEGIN columna="20" linea="16" longitud="5"/>
  <IDENTIFICADOR columna="12" linea="17" longitud="1" nombre="f"/>
  <asignacion columna="13" linea="17" longitud="2"/>
  <IDENTIFICADOR columna="15" linea="17" longitud="5" nombre="fib_1"/>
  <punto_y_coma columna="20" linea="17" longitud="1"/>
  <WRITE columna="12" linea="18" longitud="5"/>
  <IDENTIFICADOR columna="18" linea="18" longitud="1" nombre="f"/>
  <punto_y_coma columna="19" linea="18" longitud="1"/>
  <END columna="8" linea="19" longitud="3"/>
  <punto_y_coma columna="11" linea="19" longitud="1"/>
  <IF columna="8" linea="20" longitud="2"/>
  <IDENTIFICADOR columna="11" linea="20" longitud="1" nombre="n"/>
  <mayor_que columna="12" linea="20" longitud="1"/>
  <NUMERO columna="13" linea="20" longitud="1" valor="1"/>
  <THEN columna="15" linea="20" longitud="4"/>
  <BEGIN columna="20" linea="20" longitud="5"/>
  <IDENTIFICADOR columna="12" linea="21" longitud="3" nombre="f_1"/>
  <asignacion columna="15" linea="21" longitud="2"/>
  <NUMERO columna="17" linea="21" longitud="1" valor="1"/>
  <punto_y_coma columna="18" linea="21" longitud="1"/>
  <WRITE columna="12" linea="22" longitud="5"/>
  <IDENTIFICADOR columna="18" linea="22" longitud="3" nombre="f_1"/>
  <punto_y_coma columna="21" linea="22" longitud="1"/>
  <IDENTIFICADOR columna="12" linea="24" longitud="3" nombre="f_2"/>
  <asignacion columna="15" linea="24" longitud="2"/>
  <NUMERO columna="17" linea="24" longitud="1" valor="1"/>
  <punto_y_coma columna="18" linea="24" longitud="1"/>
  <WRITE columna="12" linea="25" longitud="5"/>
  <IDENTIFICADOR columna="18" linea="25" longitud="3" nombre="f_2"/>
  <punto_y_coma columna="21" linea="25" longitud="1"/>
  <IDENTIFICADOR columna="12" linea="27" longitud="1" nombre="i"/>
  <asignacion columna="13" linea="27" longitud="2"/>
  <NUMERO columna="15" linea="27" longitud="1" valor="2"/>
  <punto_y_coma columna="16" linea="27" longitud="1"/>
  <WHILE columna="12" linea="28" longitud="5"/>
  <IDENTIFICADOR columna="18" linea="28" longitud="1" nombre="i"/>
  <menor_que columna="19" linea="28" longitud="1"/>
  <IDENTIFICADOR columna="20" linea="28" longitud="1" nombre="n"/>
  <DO columna="22" linea="28" longitud="2"/>
  <BEGIN columna="25" linea="28" longitud="5"/>
  <IDENTIFICADOR columna="16" linea="29" longitud="1" nombre="f"/>
  <asignacion columna="17" linea="29" longitud="2"/>
  <IDENTIFICADOR columna="19" linea="29" longitud="3" nombre="f_1"/>
  <mas columna="22" linea="29" longitud="1"/>
  <IDENTIFICADOR columna="23" linea="29" longitud="3" nombre="f_2"/>
  <punto_y_coma columna="26" linea="29" longitud="1"/>
  <WRITE columna="16" linea="30" longitud="5"/>
  <IDENTIFICADOR columna="22" linea="30" longitud="1" nombre="f"/>
  <punto_y_coma columna="23" linea="30" longitud="1"/>
  <IDENTIFICADOR columna="16" linea="31" longitud="3" nombre="f_2"/>
  <asignacion columna="19" linea="31" longitud="2"/>
  <IDENTIFICADOR columna="21" linea="31" longitud="3" nombre="f_1"/>
  <punto_y_coma columna="24" linea="31" longitud="1"/>
  <IDENTIFICADOR columna="16" linea="32" longitud="3" nombre="f_1"/>
  <asignacion columna="19" linea="32" longitud="2"/>
  <IDENTIFICADOR columna="21" linea="32" longitud="1" nombre="f"/>
  <punto_y_coma columna="22" linea="32" longitud="1"/>
  <IDENTIFICADOR columna="16" linea="33" longitud="1" nombre="i"/>
  <asignacion columna="17" linea="33" longitud="2"/>
  <IDENTIFICADOR columna="19" linea="33" longitud="1" nombre="i"/>
  <mas columna="20" linea="33" longitud="1"/>
  <NUMERO columna="21" linea="33" longitud="1" valor="1"/>
  <punto_y_coma columna="22" linea="33" longitud="1"/>
  <END columna="12" linea="34" longitud="3"/>
  <punto_y_coma columna="15" linea="34" longitud="1"/>
  <IDENTIFICADOR columna="12" linea="35" longitud="1" nombre="f"/>
  <asignacion columna="13" linea="35" longitud="2"/>
  <IDENTIFICADOR columna="15" linea="35" longitud="3" nombre="f_1"/>
  <mas columna="18" linea="35" longitud="1"/>
  <IDENTIFICADOR columna="19" linea="35" longitud="3" nombre="f_2"/>
  <punto_y_coma columna="22" linea="35" longitud="1"/>
  <END columna="8" linea="36" longitud="3"/>
  <punto_y_coma columna="11" linea="36" longitud="1"/>
  <END columna="4" linea="37" longitud="3"/>
  <punto_y_coma columna="7" linea="37" longitud="1"/>
  <BEGIN columna="0" linea="39" longitud="5"/>
  <READ columna="4" linea="40" longitud="4"/>
  <IDENTIFICADOR columna="9" linea="40" longitud="1" nombre="n"/>
  <punto_y_coma columna="10" linea="40" longitud="1"/>
  <CALL columna="4" linea="41" longitud="4"/>
  <IDENTIFICADOR columna="9" linea="41" longitud="9" nombre="fibonacci"/>
  <punto_y_coma columna="18" linea="41" longitud="1"/>
  <WRITE columna="4" linea="42" longitud="5"/>
  <IDENTIFICADOR columna="10" linea="42" longitud="1" nombre="f"/>
  <punto_y_coma columna="11" linea="42" longitud="1"/>
  <END columna="0" linea="43" longitud="3"/>
  <punto columna="3" linea="43" longitud="1"/>
  <fuente>  <![CDATA[(*
    Cálculo de los números de la serie de fibonacci:
    f_0 = 1
    f_1 = 1
    f_n = f_{n-1} + f_{n-2}, n>1
*)
const fib_0=1, fib_1=1;
var n, f;

procedure fibonacci;
    var i,   (* Variable para hacer el recorrido *)
        f_1, (* Número Fibonacci anterior *)
        f_2; (* Número Fibonacci anterior al anterior *)
    begin
        if n=0 then f:=fib_0;
        if n=1 then begin
            f:=fib_1;
            write f; (* Los primeros dos elementos son iguales *)
        end;
        if n>1 then begin
            f_1:=1;
            write f_1;
            
            f_2:=1;
            write f_2;
            
            i:=2;
            while i<n do begin
                f:=f_1+f_2;
                write f;
                f_2:=f_1;
                f_1:=f;
                i:=i+1;
            end;
            f:=f_1+f_2;
        end;
    end;

begin
    read n;
    call fibonacci;
    write f;
end.
]]>  </fuente>
</lexemas>
\end{lstlisting}

\subsection{Descripción de la Fase de Análisis Sintáctico}

El propósito de esta fase es convertir ---y al mismo tiempo verificar
si es posible convertir--- la secuencia de elementos léxicos proporcionada
por la fase de análisis léxico, en un árbol de sintaxis que representa
una cadena que puede ser generada por la gramática del lenguaje fuente.\cite{Compiler Construction,Introducci=0000F3n a los compiladores,Libro del Drag=0000F3n}

La fase de análisis sintáctico genera el árbol de sintaxis correspondiente
al programa fuente. La estructura general del árbol de sintaxis para
todo programa \textsf{pl0+} es la siguiente:

\begin{lstlisting}
<arbol_de_sintaxis>
  <programa>
    <bloque>
      ...
    </bloque>
  </programa>
  <fuente>  <![CDATA[...]]>  </fuente>
</arbol_de_sintaxis>
\end{lstlisting}

De acuerdo a la sintaxis de \textsf{pl0+}, todo programa está compuesto
por un bloque principal de código.

Todo bloque se representa de la siguiente manera (una secuencia de
declaraciones de constantes, variables y procedimientos y opcionalmente
una instrucción):

\begin{lstlisting}
<bloque>
  <constante columna="15" linea="7" nombre="fib_1" valor="1"/>
  .
  .
  .
  <variable columna="4" linea="8" nombre="n"/>
  .
  .
  .
  <procedimiento columna="10" linea="10" nombre="fibonacci">
    <bloque> ... </bloque>
  </procedimiento>
  .
  .
  .
  <!-- Una instrucción opcional -->
</bloque>
\end{lstlisting}

Las instrucciones de asignación, llamada a procedimiento, secuencias
\texttt{begin}/\texttt{end}, condicionales \texttt{if}, ciclos \texttt{while}
y operaciones de lectura y escritura se representan así:

\begin{lstlisting}
<!-- n := ... -->
<asignacion variable="n">
  <!-- una expresión -->
</asignacion>

<!-- call fibonacci -->
<llamada procedimiento="fibonacci"/>

<!-- begin ... end -->
<secuencia>
  <!-- una o más instrucciones -->
</secuencia>

<!-- if ... then ... else ... -->
<condicional>
  <condicion operacion="...">
    <!-- una o dos expresiones operandos -->
  </condicion>
  <!-- instrucción para verdadero -->
  <!-- instrucción opcional para falso -->
</condicional>

<!-- while ... do ... -->
<ciclo>
  <condicion operacion="...">
    <!-- una o dos expresiones operandos -->
  </condicion>
  <!-- instrucción a repetir -->
</ciclo>

<!-- read n -->
<leer variable="n"/>

<!-- write f -->
<escribir simbolo="f"/>
\end{lstlisting}

Los parámetros válidos para el atributo \texttt{operacion} de la etiqueta
\texttt{condicion} son: \texttt{comparacion} (\texttt{=}), \texttt{diferente\_de}
(\texttt{<>}), \texttt{menor\_que} (\texttt{<}), \texttt{mayor\_que}
(\texttt{>}), \texttt{menor\_igual\_que} (\texttt{<=}), \texttt{mayor\_igual\_que}
(\texttt{>=}).

Las etiquetas válidas como expresión son \texttt{suma}, \texttt{resta},
\texttt{multiplicacion}, \texttt{division}, \texttt{negativo} y las
etiquetas especiales: \lstinline!<identificador simbolo="f_1"/>!
y \lstinline!<numero valor="10"/>! para identificadores y literales
respectivamente.

A continuación se presenta un ejemplo de programa sencillo ---válido
pero no funcional--- con una expresión y un ciclo (programa \ref{expresion.pl0+})
y la salida correspondiente de su análisis sintáctico (programa \ref{expresion.pl0+sin}):

\begin{lstlisting}[language=Pascal,caption={Programa con una expresión y un ciclo},label={expresion.pl0+}]
const n=5;
var f, i;
begin
    i := (f*-2)*(n*i) + 5*-9/8*i;
    write i;
    while f>i do write f;
end.
\end{lstlisting}

\begin{lstlisting}[caption={Salida del Análisis Sintáctico del programa enterior},label={expresion.pl0+sin}]
<?xml version="1.0" ?>
<arbol_de_sintaxis>
  <!--
**************************************
Por favor, no modifique este archivo
a menos que sepa lo que está haciendo
**************************************
-->
<programa>
<bloque>
  <constante columna="6" linea="1" nombre="n" valor="5"/>
  <variable columna="4" linea="2" nombre="f"/>
  <variable columna="7" linea="2" nombre="i"/>
  <secuencia columna="0" linea="3">
    <asignacion columna="4" linea="4" variable="i">
      <suma columna="22" linea="4">
        <multiplicacion columna="15" linea="4">
          <multiplicacion columna="11" linea="4">
            <identificador columna="10" linea="4" simbolo="f"/>
            <negativo columna="12" linea="4">
              <numero columna="13" linea="4" valor="2"/>
            </negativo>
          </multiplicacion>
          <multiplicacion columna="18" linea="4">
            <identificador columna="17" linea="4" simbolo="n"/>
            <identificador columna="19" linea="4" simbolo="i"/>
          </multiplicacion>
        </multiplicacion>
        <multiplicacion columna="30" linea="4">
          <multiplicacion columna="25" linea="4">
            <numero columna="24" linea="4" valor="5"/>
            <division columna="28" linea="4">
              <negativo columna="26" linea="4">
                <numero columna="27" linea="4" valor="9"/>
              </negativo>
              <numero columna="29" linea="4" valor="8"/>
            </division>
          </multiplicacion>
          <identificador columna="31" linea="4" simbolo="i"/>
        </multiplicacion>
      </suma>
    </asignacion>
    <escribir columna="10" linea="5" simbolo="i"/>
    <ciclo columna="4" linea="6">
      <condicion columna="10" linea="6" operacion="mayor_que">
        <identificador columna="10" linea="6" simbolo="f"/>
        <identificador columna="12" linea="6" simbolo="i"/>
      </condicion>
      <escribir columna="23" linea="6" simbolo="f"/>
    </ciclo>
  </secuencia>
</bloque>
</programa>
  <fuente>
<![CDATA[const n=5;
var f, i;
begin
    i := (f*-2)*(n*i) + 5*-9/8*i;
    write i;
    while f>i do write f;
end.
]]>  </fuente>
</arbol_de_sintaxis>
\end{lstlisting}

\subsection{Descripción de la Fase de Análisis Semántico}

El propósito de esta fase es tomar el árbol de sintaxis y realizar
las siguientes actividades:\cite{Introducci=0000F3n a los compiladores,Libro del Drag=0000F3n}
\begin{enumerate}
\item Revisión de la Coherencia de tipos de dato.\\
Por ejemplo, verificar que a una variable de tipo cadena no se le
asigne un número entero (suponiendo que esto no sea válido en el lenguaje
fuente).
\item Conversión implícita entre tipos de dato.\\
Por ejemplo, cuando una variable de tipo numérico flotante se pasa
como parámetro real a una función que recibe un número entero como
parámetro formal, esta fase debe hacer la conversión implícita de
flotante a entero si esta aplica o reportar el error.
\item Comprobación del flujo de control.\\
Por ejemplo, las proposiciones \texttt{break} y \texttt{continue}
de muchos lenguajes de programación transfieren el flujo de control
de un punto a otro. Esta fase debe verificar si existe algún punto
válido a dónde transferir el flujo de control (un \texttt{for}, \texttt{while},
\texttt{switch}, \texttt{case}, etc.).
\item Comprobaciones de unicidad.\\
Por ejemplo verificar que las opciones de un bloque \texttt{case}
o \texttt{switch} no estén repetidas, o que un identificador no esté
declarado más de una vez en un mismo ámbito.
\item Comprobaciones relacionadas con nombres.\\
Por ejemplo en los lenguajes en los que la palabra \texttt{end} (o
equivalente) va seguida de un identificador que debe corresponder
con el identificador al inicio del bloque.
\end{enumerate}
En el caso del compilador \texttt{tradukilo.py}, las funciones de
esta fase son:
\begin{enumerate}
\item Colocar un código único a cada símbolo (de variable, de constante
y de procedimiento) para ser referenciado después. El código incluye
información del tipo de símbolo y el ámbito en el que está declarado.
\item Verificar duplicidad de símbolos en el mismo ámbito.
\item Verificar si hay identificadores de procedimiento referenciados en
una expresión o en una asignacion (lo cual no es válido en \textsf{pl0+}).
\item Verificar si hay identificadores de constante referenciados en el
lado izquierdo de una asignación.
\item Verificar que cada identificador/símbolo referenciado esté en un ámbito
válido. Es decir, verifica la ``integridad referencial'' de los
símbolos.
\end{enumerate}
Los códigos asignados en esta fase a los programas \texttt{pl0+} siguen
las siguientes reglas:
\begin{enumerate}
\item El código de todo bloque tiene el prefijo ``\texttt{b}''.
\item El código de toda constante tiene el prefijo ``\texttt{c}''.
\item El código de toda variable tiene el prefijo ``\texttt{v}''.
\item El bloque principal siempre tiene el código ``\texttt{b0}''.
\item Todos los códigos (excepto el del bloque principal) tienen un prefijo
formado por el caracter ``\texttt{\_}'' y un número correlativo
---que comienza desde cero--- para el tipo de declaración dentro
de ese ámbito.\\
Por ejemplo, el código ``\texttt{v0/2/1\_3}'' indica que es la cuarta
variable del bloque donde está declarada; ``\texttt{c0\_0}'' indica
que es la primera constante del bloque donde está declarada; ``\texttt{b0/2\_1}''
indica que es el segundo bloque declarado en su ámbito; etc.
\item El cuerpo de los códigos indica la ruta de anidamiento en el que los
símbolos están declarados.
\end{enumerate}
Para ilustrar la semántica de los códigos de los identificadores,
consideremos el siguiente programa fuente:

\begin{lstlisting}[language=Pascal,caption={codigos.pl0+ -- Ejemplo de análisis semántico},label={codigos.pl0+}]
const const_global=56, otra_const_global=9;
var var_global, otra_var_global;
procedure proc;
    write const_global;

procedure proc2;
    call proc;

procedure otro;
    const g=9, m=5;
    var temp, temp2;
    procedure otro2;
        procedure vacio;;
        procedure otro_vacio;; (*Procedimientos vacíos*)
        procedure otro3;
            var i,otro3_var;
            begin
                write g;
                write otra_const_global;
                write m;
                write temp2;
                call proc2;
            end
        ;
    ;
;

write const_global;
.
\end{lstlisting}

Por ejemplo, analicemos el código de la variable \texttt{otro3\_var}
en el programa \ref{codigos.pl0+}, que es ``\texttt{v0/2/0/2\_1}'',
lo que puede verse en el programa \ref{codigos.pl0+sem}. Este código
indica que el elemento es una variable, y es la segunda de su ámbito
(la primera es \texttt{i}).

El cuerpo del código ---la secuencia ``\texttt{0/2/0/2}'' del código
de la variable--- indica que está declarada en el bloque con código
``\texttt{b0/2/0\_2}'' que es \texttt{otro3} según vemos en el programa
\ref{codigos.pl0+sem}. El último correlativo del cuerpo del código
de la variable (que es ``\texttt{2}'') indica el correlativo de
declaración del bloque en el que está declarada, y el resto del cuerpo
sin el caracter ``\texttt{/}'' (que es ``\texttt{0/2/0}'') indica
el cuerpo del código del bloque en el que está declarada. De ahí se
sabe que la variable \texttt{otro3\_var} está declarada en el bloque
\texttt{otro3}.

El cuerpo completo del código de la variable también permite rastrear
toda la jerarquía de procedimientos en los que está inserta. La siguiente
tabla lo ilustra:

\begin{tabular}{|c|c|c|}
\hline 
Elemento & Código & Contenedor inmediato\tabularnewline
\hline 
\hline 
\texttt{otro3\_var} & \texttt{v0/2/0/2\_1} & \texttt{otro3}\tabularnewline
\hline 
\texttt{otro3} & \texttt{b0/2/0\_2} & \texttt{otro2}\tabularnewline
\hline 
\texttt{otro2} & \texttt{b0/2\_0} & \texttt{otro}\tabularnewline
\hline 
\texttt{otro} & \texttt{b0\_2} & bloque principal\tabularnewline
\hline 
bloque principal & \texttt{b0} & --\tabularnewline
\hline 
\end{tabular}

A continuación sigue el resultado del análisis semántico del programa
\ref{codigos.pl0+}:

\begin{lstlisting}[caption={codigos.pl0+sem -- Análisis Semántico del programa anterior},label={codigos.pl0+sem}]
<?xml version="1.0" ?>
<arbol_de_sintaxis_revisado>
  <!--
**************************************
Por favor, no modifique este archivo
a menos que sepa lo que está haciendo
**************************************
-->
<programa>
<bloque codigo="b0">
  <constante codigo="c0_0" columna="6" linea="1" nombre="const_global" valor="56"/>
  <constante codigo="c0_1" columna="23" linea="1" nombre="otra_const_global" valor="9"/>
  <variable codigo="v0_0" columna="4" linea="2" nombre="var_global"/>
  <variable codigo="v0_1" columna="16" linea="2" nombre="otra_var_global"/>
  <procedimiento columna="10" linea="3" nombre="proc">
    <bloque codigo="b0_0">
      <escribir codigo="c0_0" columna="10" linea="4" simbolo="const_global" valor="56"/>
    </bloque>
  </procedimiento>
  <procedimiento columna="10" linea="6" nombre="proc2">
    <bloque codigo="b0_1">
      <llamada codigo="b0_0" columna="9" linea="7" procedimiento="proc"/>
    </bloque>
  </procedimiento>
  <procedimiento columna="10" linea="9" nombre="otro">
    <bloque codigo="b0_2">
      <constante codigo="c0/2_0" columna="10" linea="10" nombre="g" valor="9"/>
      <constante codigo="c0/2_1" columna="15" linea="10" nombre="m" valor="5"/>
      <variable codigo="v0/2_0" columna="8" linea="11" nombre="temp"/>
      <variable codigo="v0/2_1" columna="14" linea="11" nombre="temp2"/>
      <procedimiento columna="14" linea="12" nombre="otro2">
        <bloque codigo="b0/2_0">
          <procedimiento columna="18" linea="13" nombre="vacio">
            <bloque codigo="b0/2/0_0"/>
          </procedimiento>
          <procedimiento columna="18" linea="14" nombre="otro_vacio">
            <bloque codigo="b0/2/0_1"/>
          </procedimiento>
          <procedimiento columna="18" linea="15" nombre="otro3">
            <bloque codigo="b0/2/0_2">
              <variable codigo="v0/2/0/2_0" columna="16" linea="16" nombre="i"/>
              <variable codigo="v0/2/0/2_1" columna="18" linea="16" nombre="otro3_var"/>
              <secuencia columna="12" linea="17">
                <escribir codigo="c0/2_0" columna="22" linea="18" simbolo="g" valor="9"/>
                <escribir codigo="c0_1" columna="22" linea="19" simbolo="otra_const_global" valor="9"/>
                <escribir codigo="c0/2_1" columna="22" linea="20" simbolo="m" valor="5"/>
                <escribir codigo="v0/2_1" columna="22" linea="21" simbolo="temp2"/>
                <llamada codigo="b0_1" columna="21" linea="22" procedimiento="proc2"/>
              </secuencia>
            </bloque>
          </procedimiento>
        </bloque>
      </procedimiento>
    </bloque>
  </procedimiento>
  <escribir codigo="c0_0" columna="6" linea="28" simbolo="const_global" valor="56"/>
</bloque>
</programa>
  <fuente>
<![CDATA[const const_global=56, otra_const_global=9;
var var_global, otra_var_global;
procedure proc;
    write const_global;

procedure proc2;
    call proc;

procedure otro;
    const g=9, m=5;
    var temp, temp2;
    procedure otro2;
        procedure vacio;;
        procedure otro_vacio;; (*Procedimientos vacíos*)
        procedure otro3;
            var i,otro3_var;
            begin
                write g;
                write otra_const_global;
                write m;
                write temp2;
                call proc2;
            end
        ;
    ;
;

write const_global;
.
]]>  </fuente>
</arbol_de_sintaxis_revisado>
\end{lstlisting}

\subsection{Descripción de la Fase de Generación de Código Objeto}

La generación de código objeto (lenguaje \textsf{p+}) se realiza con
el objetivo de transformar el código fuente original en ``código
máquina'' para una máquina virtual que interprete exclusivamente
programas en lenguaje \textsf{p+}.

A continuación se presentan las reglas generales de traducción al
lenguaje \textsf{p+}, y para simplificar la lectura, se define la
función $gen:C_{pl0+}\longrightarrow C_{p+}$ que hipotéticamente
convierte un fragmento de código fuente en lenguaje \textsf{pl0+}
a una serie de instrucciones en lenguaje \textsf{p+}.

\subsubsection{Procedimientos\label{subsec:Procedimientos}}

Se traducen en la siguiente secuencia de instrucciones:

\begin{center}
\begin{tabular}{rl}
$i$: & \texttt{SAL $b$}\tabularnewline
 & \texttt{$\vdots$}\tabularnewline
 & (instrucciones de los subprocedimientos)\tabularnewline
 & \texttt{$\vdots$}\tabularnewline
$b$: & \texttt{INS $num$}\tabularnewline
 & \texttt{$\vdots$}\tabularnewline
 & (instrucciones del procedimiento)\tabularnewline
 & \texttt{$\vdots$}\tabularnewline
$f$: & \texttt{RET}\tabularnewline
\end{tabular}
\par\end{center}

donde $i$, $b$ y $f$ son las direcciones de las respectivas instrucciones,
$i<b<f$, $i$ es la dirección de inicio del procedimiento y siempre
es un salto incondicional a la instrucción de instanciación del bloque,
que está en la dirección $b$, $f$ es la dirección del fin del bloque
o procedimiento, $num$ es el número de variables locales al bloque
más el espacio necesario para los registros dinámicos (que en el caso
de \textsf{p+} son 3).

En el caso del bloque principal, se cumple que $i=0$ y $f$ es la
última instrucción del programa.

\subsubsection{Asignaciones}

Las instrucciones del tipo \texttt{<var>~:=~<expresión>;} se traducen
en la siguiente secuencia de instrucciones:

\begin{center}
\begin{tabular}{rl}
 & \texttt{$gen($<expresión>$)$}\tabularnewline
 & \texttt{ALM $dif$ $pos$}\tabularnewline
\end{tabular}
\par\end{center}

Primero se genera la secuencia de instrucciones que evalúan la expresión
y luego se almacena el resultado ---que quedó en el tope de la pila---,
en la celda de memoria reservado para la variable \texttt{<var>}.

\subsubsection{Condicionales \texttt{if}}

Se traducen de manera diferente si tienen o no tienen parte \texttt{else}.
A continuación se presentan ambas formas:

\paragraph{Sin \texttt{else}}

La forma ``\texttt{if $C$ then $B_{1}$}'' se traduce como:

\begin{center}
\begin{tabular}{rl}
 & \texttt{$gen(C)$}\tabularnewline
 & \texttt{SAC $d_{1}$}\tabularnewline
 & \texttt{$gen(B_{1})$}\tabularnewline
$d_{1}$: & \texttt{$\vdots$}\tabularnewline
 & (lo que sigue al \texttt{if})\tabularnewline
\end{tabular}
\par\end{center}

Primero se genera la secuencia de instrucciones para la condición
$C$, luego se coloca un salto condicional hacia las instrucciones
que le siguen al \texttt{if}, y luego se genera la secuencia de instrucciones
para $B_{1}$.

\paragraph{Con \texttt{else}}

La forma ``\texttt{if $C$ then $B_{1}$ else }$B_{2}$'' se traduce
como:

\begin{center}
\begin{tabular}{rl}
 & \texttt{$gen(C)$}\tabularnewline
 & \texttt{SAC $d_{1}$}\tabularnewline
 & \texttt{$gen(B_{1})$}\tabularnewline
 & \texttt{SAL $d_{2}$}\tabularnewline
$d_{1}$: & \texttt{$gen(B_{2})$}\tabularnewline
$d_{2}$: & \texttt{$\vdots$}\tabularnewline
 & (lo que sigue al \texttt{if})\tabularnewline
\end{tabular}
\par\end{center}

Primero se genera la secuencia de instrucciones para la condición
$C$, luego se coloca un salto condicional hacia el inicio de las
instrucciones de $B_{2}$ (la parte del \texttt{else}), luego se genera
la secuencia de instrucciones de $B_{1}$ (la parte del \texttt{then})
y después un salto incondicional hacia las instrucciones que le siguen
al \texttt{if}. Finalmente se generan las instrucciones para $B_{2}$
(la parte \texttt{else}).

\subsubsection{Ciclos \texttt{while}}

Tienen la forma ``\texttt{while $C$ do }$B$'' y se traducen en
la siguiente secuencia de instrucciones:

\begin{center}
\begin{tabular}{rl}
$d_{1}$: & \texttt{$gen(C)$}\tabularnewline
 & \texttt{SAC $d_{2}$}\tabularnewline
 & \texttt{$gen(B)$}\tabularnewline
 & \texttt{SAL $d_{1}$}\tabularnewline
$d_{2}$: & $\vdots$\tabularnewline
 & (lo que sigue al \texttt{while})\tabularnewline
\end{tabular}
\par\end{center}

Primero se genera la secuencia de instrucciones para la condición
$C$, luego se coloca un salto condicional hacia la instrucción que
le sigue al ciclo \texttt{while}, luego se genera la secuencia de
instrucciones del cuerpo del ciclo y luego se coloca un salto incondicional
hacia el inicio de las instrucciones de la condición $C$.

\subsubsection{Llamadas a procedimiento}

Tienen la forma ``\texttt{call~<proc>}'' y se traducen de la siguiente
manera:

\begin{center}
\begin{tabular}{rl}
 & \texttt{$\vdots$}\tabularnewline
$b$: & \texttt{INS $n$}\tabularnewline
 & \texttt{$\vdots$}\tabularnewline
 & (instrucciones del procedimiento \texttt{<proc>})\tabularnewline
 & \texttt{$\vdots$}\tabularnewline
$f$: & \texttt{RET}\tabularnewline
 & \texttt{$\vdots$}\tabularnewline
$d$: & \texttt{LLA $dif$ $b$}\tabularnewline
$d+1$: & \texttt{$\vdots$}\tabularnewline
 & (lo que sigue a la llamada)\tabularnewline
\end{tabular}
\par\end{center}

La instrucción en la dirección $b$ marca el inicio operativo del
procedimiento invocado ---que siempre está constituido por una instrucción
\texttt{INS}--- y $f$ su fin ---que siempre está consituido por
una instrucción \texttt{RET}--- (ver sección \ref{subsec:Procedimientos}).
Cuando se ejecuta una instrucción como la de la dirección $d$, se
guardan los registros dinámicos necesarios y se salta a la dirección
$b$. El valor $dif$ sirve para calcular el enlace estático para
poder identificar las variables de los procedimientos de ámbito superior.
Cuando se llega a la instrucción en la dirección $f$, se regresa
el control a la dirección $d+1$ y se reestablecen los registros dinámicos
del bloque anterior.

\subsubsection{Escritura y Lectura}

Las instrucciones de la forma ``\texttt{read~<var>}'' leen un entero
de la entrada estándar y la almacenan en la variable \texttt{<var>}.
Las instrucciones de la forma ``\texttt{write~<var>}'' escriben
el valor de la variable \texttt{<var>} en la salida estándar. Se traducen
a las siguientes secuencias:

\begin{center}
\begin{tabular}{|l|l|l|}
\cline{1-1} \cline{3-3} 
\texttt{read~<var>:} &  & \texttt{write~<var>:}\tabularnewline
 &  & \tabularnewline
\texttt{LEE} &  & \texttt{CAR $dif$ $pos$}\tabularnewline
\texttt{ALM $dif$ $pos$} &  & \texttt{ESC}\tabularnewline
\cline{1-1} \cline{3-3} 
\end{tabular}
\par\end{center}

\subsubsection{Expresiones}

Un fragmento de código \textsf{pl0+} del tipo ``\texttt{$operando_{1}$~$operador$~$operando_{2}$}''
se traducirá como sigue:

\begin{center}
\begin{tabular}{ll}
 & $gen(operando_{1})$\tabularnewline
 & $gen(operando_{2})$\tabularnewline
 & \texttt{OPR $c\acute{o}d$}\tabularnewline
\end{tabular}
\par\end{center}

donde $c\acute{o}d$ es el código de operación de $operador$. Los
códigos de operación son los mostrados en la siguiente tabla:

\begin{center}
\begin{tabular}{|c|c|}
\hline 
{\footnotesize{}Operación} & {\footnotesize{}Código}\tabularnewline
\hline 
\hline 
\texttt{\footnotesize{}negativo} & {\footnotesize{}1}\tabularnewline
\hline 
\texttt{\footnotesize{}suma} & {\footnotesize{}2}\tabularnewline
\hline 
\texttt{\footnotesize{}resta} & {\footnotesize{}3}\tabularnewline
\hline 
\texttt{\footnotesize{}multiplicacion} & {\footnotesize{}4}\tabularnewline
\hline 
\texttt{\footnotesize{}division} & {\footnotesize{}5}\tabularnewline
\hline 
\texttt{\footnotesize{}odd}{\footnotesize{} (impar)} & {\footnotesize{}6}\tabularnewline
\hline 
\texttt{\footnotesize{}comparacion} & {\footnotesize{}8}\tabularnewline
\hline 
\texttt{\footnotesize{}diferente\_de} & {\footnotesize{}9}\tabularnewline
\hline 
\texttt{\footnotesize{}menor\_que} & {\footnotesize{}10}\tabularnewline
\hline 
\texttt{\footnotesize{}mayor\_igual\_que} & {\footnotesize{}11}\tabularnewline
\hline 
\texttt{\footnotesize{}mayor\_que} & {\footnotesize{}12}\tabularnewline
\hline 
\texttt{\footnotesize{}menor\_igual\_que} & {\footnotesize{}13}\tabularnewline
\hline 
\end{tabular}
\par\end{center}

\subsubsection{Ejemplo}

A continuación se presenta el resultado de la fase de generación de
código del programa \vref{fibonacci.pl0+}:

\begin{lstlisting}[caption={fibonacci.p+ -- Código objeto del programa fibonacci.pl0+},label={fibonacci.p+}]
<?xml version="1.0" ?>
<codigo_pmas>
  <!--
**************************************
Por favor, no modifique este archivo
a menos que sepa lo que está haciendo
**************************************
-->
<salto_incondicional direccion="0" parametro="55">
  <informacion codigo="b0" inicio_de_procedimiento="--PRINCIPAL--"/>
</salto_incondicional>
<salto_incondicional direccion="1" parametro="2">
  <informacion codigo="b0_0" columna="10" inicio_de_procedimiento="fibonacci" linea="10"/>
</salto_incondicional>
<instanciar_procedimiento direccion="2" parametro="6">
  <informacion codigo="b0_0" columna="10" inicio_de_procedimiento="fibonacci" linea="10"/>
</instanciar_procedimiento>
<cargar_variable diffnivel="1" direccion="3" parametro="3">
  <informacion columna="8" linea="15">
    Inicio de condicional (if-then)
  </informacion>
  <informacion codigo="v0_0" columna="11" linea="15" variable="n"/>
</cargar_variable>
<literal direccion="4" parametro="0">
  <informacion columna="13" linea="15"/>
</literal>
<operacion direccion="5" parametro="8">
  <informacion columna="11" linea="15" operacion="comparacion"/>
</operacion>
<salto_condicional direccion="6" parametro="9">
  <informacion columna="8" linea="15">
    Inicio del cuerpo del 'then'
  </informacion>
</salto_condicional>
<literal direccion="7" parametro="1">
  <informacion codigo="v0_1" columna="20" linea="15" variable="f">
    Inicio de asignación de variable
  </informacion>
  <informacion codigo="c0_0" columna="23" constante="fib_0" linea="15"/>
</literal>
<almacenar_variable diffnivel="1" direccion="8" parametro="4">
  <informacion codigo="v0_1" columna="20" linea="15" variable="f"/>
  <informacion columna="8" linea="15">
    Fin de condicional (if-then)
  </informacion>
</almacenar_variable>
<cargar_variable diffnivel="1" direccion="9" parametro="3">
  <informacion columna="8" linea="16">
    Inicio de condicional (if-then)
  </informacion>
  <informacion codigo="v0_0" columna="11" linea="16" variable="n"/>
</cargar_variable>
<literal direccion="10" parametro="1">
  <informacion columna="13" linea="16"/>
</literal>
<operacion direccion="11" parametro="8">
  <informacion columna="11" linea="16" operacion="comparacion"/>
</operacion>
<salto_condicional direccion="12" parametro="17">
  <informacion columna="8" linea="16">
    Inicio del cuerpo del 'then'
  </informacion>
</salto_condicional>
<literal direccion="13" parametro="1">
  <informacion codigo="v0_1" columna="12" linea="17" variable="f">
    Inicio de asignación de variable
  </informacion>
  <informacion codigo="c0_1" columna="15" constante="fib_1" linea="17"/>
</literal>
<almacenar_variable diffnivel="1" direccion="14" parametro="4">
  <informacion codigo="v0_1" columna="12" linea="17" variable="f"/>
</almacenar_variable>
<cargar_variable diffnivel="1" direccion="15" parametro="4">
  <informacion codigo="v0_1" columna="18" linea="18" variable="f"/>
</cargar_variable>
<escribir direccion="16">
  <informacion columna="8" linea="16">
    Fin de condicional (if-then)
  </informacion>
</escribir>
<cargar_variable diffnivel="1" direccion="17" parametro="3">
  <informacion columna="8" linea="20">
    Inicio de condicional (if-then)
  </informacion>
  <informacion codigo="v0_0" columna="11" linea="20" variable="n"/>
</cargar_variable>
<literal direccion="18" parametro="1">
  <informacion columna="13" linea="20"/>
</literal>
<operacion direccion="19" parametro="12">
  <informacion columna="11" linea="20" operacion="mayor_que"/>
</operacion>
<salto_condicional direccion="20" parametro="54">
  <informacion columna="8" linea="20">
    Inicio del cuerpo del 'then'
  </informacion>
</salto_condicional>
<literal direccion="21" parametro="1">
  <informacion codigo="v0/0_1" columna="12" linea="21" variable="f_1">
    Inicio de asignación de variable
  </informacion>
  <informacion columna="17" linea="21"/>
</literal>
<almacenar_variable diffnivel="0" direccion="22" parametro="4">
  <informacion codigo="v0/0_1" columna="12" linea="21" variable="f_1"/>
</almacenar_variable>
<cargar_variable diffnivel="0" direccion="23" parametro="4">
  <informacion codigo="v0/0_1" columna="18" linea="22" variable="f_1"/>
</cargar_variable>
<escribir direccion="24"/>
<literal direccion="25" parametro="1">
  <informacion codigo="v0/0_2" columna="12" linea="24" variable="f_2">
    Inicio de asignación de variable
  </informacion>
  <informacion columna="17" linea="24"/>
</literal>
<almacenar_variable diffnivel="0" direccion="26" parametro="5">
  <informacion codigo="v0/0_2" columna="12" linea="24" variable="f_2"/>
</almacenar_variable>
<cargar_variable diffnivel="0" direccion="27" parametro="5">
  <informacion codigo="v0/0_2" columna="18" linea="25" variable="f_2"/>
</cargar_variable>
<escribir direccion="28"/>
<literal direccion="29" parametro="2">
  <informacion codigo="v0/0_0" columna="12" linea="27" variable="i">
    Inicio de asignación de variable
  </informacion>
  <informacion columna="15" linea="27"/>
</literal>
<almacenar_variable diffnivel="0" direccion="30" parametro="3">
  <informacion codigo="v0/0_0" columna="12" linea="27" variable="i"/>
</almacenar_variable>
<cargar_variable diffnivel="0" direccion="31" parametro="3">
  <informacion columna="12" linea="28">
    Inicio de ciclo (while)
  </informacion>
  <informacion codigo="v0/0_0" columna="18" linea="28" variable="i"/>
</cargar_variable>
<cargar_variable diffnivel="1" direccion="32" parametro="3">
  <informacion codigo="v0_0" columna="20" linea="28" variable="n"/>
</cargar_variable>
<operacion direccion="33" parametro="10">
  <informacion columna="18" linea="28" operacion="menor_que"/>
</operacion>
<salto_condicional direccion="34" parametro="50">
  <informacion columna="12" linea="28">
    Inicio del cuerpo del ciclo
  </informacion>
</salto_condicional>
<cargar_variable diffnivel="0" direccion="35" parametro="4">
  <informacion codigo="v0_1" columna="16" linea="29" variable="f">
    Inicio de asignación de variable
  </informacion>
  <informacion codigo="v0/0_1" columna="19" linea="29" variable="f_1"/>
</cargar_variable>
<cargar_variable diffnivel="0" direccion="36" parametro="5">
  <informacion codigo="v0/0_2" columna="23" linea="29" variable="f_2"/>
</cargar_variable>
<operacion direccion="37" parametro="2">
  <informacion columna="22" linea="29" operacion="suma"/>
</operacion>
<almacenar_variable diffnivel="1" direccion="38" parametro="4">
  <informacion codigo="v0_1" columna="16" linea="29" variable="f"/>
</almacenar_variable>
<cargar_variable diffnivel="1" direccion="39" parametro="4">
  <informacion codigo="v0_1" columna="22" linea="30" variable="f"/>
</cargar_variable>
<escribir direccion="40"/>
<cargar_variable diffnivel="0" direccion="41" parametro="4">
  <informacion codigo="v0/0_2" columna="16" linea="31" variable="f_2">
    Inicio de asignación de variable
  </informacion>
  <informacion codigo="v0/0_1" columna="21" linea="31" variable="f_1"/>
</cargar_variable>
<almacenar_variable diffnivel="0" direccion="42" parametro="5">
  <informacion codigo="v0/0_2" columna="16" linea="31" variable="f_2"/>
</almacenar_variable>
<cargar_variable diffnivel="1" direccion="43" parametro="4">
  <informacion codigo="v0/0_1" columna="16" linea="32" variable="f_1">
    Inicio de asignación de variable
  </informacion>
  <informacion codigo="v0_1" columna="21" linea="32" variable="f"/>
</cargar_variable>
<almacenar_variable diffnivel="0" direccion="44" parametro="4">
  <informacion codigo="v0/0_1" columna="16" linea="32" variable="f_1"/>
</almacenar_variable>
<cargar_variable diffnivel="0" direccion="45" parametro="3">
  <informacion codigo="v0/0_0" columna="16" linea="33" variable="i">
    Inicio de asignación de variable
  </informacion>
  <informacion codigo="v0/0_0" columna="19" linea="33" variable="i"/>
</cargar_variable>
<literal direccion="46" parametro="1">
  <informacion columna="21" linea="33"/>
</literal>
<operacion direccion="47" parametro="2">
  <informacion columna="20" linea="33" operacion="suma"/>
</operacion>
<almacenar_variable diffnivel="0" direccion="48" parametro="3">
  <informacion codigo="v0/0_0" columna="16" linea="33" variable="i"/>
</almacenar_variable>
<salto_incondicional direccion="49" parametro="31">
  <informacion columna="12" linea="28">
    Fin del cuerpo del ciclo
  </informacion>
</salto_incondicional>
<cargar_variable diffnivel="0" direccion="50" parametro="4">
  <informacion codigo="v0_1" columna="12" linea="35" variable="f">
    Inicio de asignación de variable
  </informacion>
  <informacion codigo="v0/0_1" columna="15" linea="35" variable="f_1"/>
</cargar_variable>
<cargar_variable diffnivel="0" direccion="51" parametro="5">
  <informacion codigo="v0/0_2" columna="19" linea="35" variable="f_2"/>
</cargar_variable>
<operacion direccion="52" parametro="2">
  <informacion columna="18" linea="35" operacion="suma"/>
</operacion>
<almacenar_variable diffnivel="1" direccion="53" parametro="4">
  <informacion codigo="v0_1" columna="12" linea="35" variable="f"/>
  <informacion columna="8" linea="20">
    Fin de condicional (if-then)
  </informacion>
</almacenar_variable>
<retornar direccion="54">
  <informacion fin_de_procedimiento="fibonacci"/>
</retornar>
<instanciar_procedimiento direccion="55" parametro="5">
  <informacion codigo="b0" inicio_de_procedimiento="--PRINCIPAL--"/>
</instanciar_procedimiento>
<leer direccion="56">
  <informacion codigo="v0_0" columna="9" linea="40" variable="n"/>
</leer>
<almacenar_variable diffnivel="0" direccion="57" parametro="3">
  <informacion codigo="v0_0" columna="9" linea="40" variable="n"/>
</almacenar_variable>
<llamar_procedimiento diffnivel="0" direccion="58" parametro="1">
  <informacion codigo="b0_0" columna="9" linea="41" procedimiento="fibonacci"/>
</llamar_procedimiento>
<cargar_variable diffnivel="0" direccion="59" parametro="4">
  <informacion codigo="v0_1" columna="10" linea="42" variable="f"/>
</cargar_variable>
<escribir direccion="60"/>
<retornar direccion="61">
  <informacion fin_de_procedimiento="--PRINCIPAL--"/>
</retornar>
  <ensamblador>
<![CDATA[
   0 SAL    -         55
   1 SAL    -          2
   2 INS    -          6
   3 CAR    1          3
   4 LIT    -          0
   5 OPR    -          8
   6 SAC    -          9
   7 LIT    -          1
   8 ALM    1          4
   9 CAR    1          3
  10 LIT    -          1
  11 OPR    -          8
  12 SAC    -         17
  13 LIT    -          1
  14 ALM    1          4
  15 CAR    1          4
  16 ESC    -          -
  17 CAR    1          3
  18 LIT    -          1
  19 OPR    -         12
  20 SAC    -         54
  21 LIT    -          1
  22 ALM    0          4
  23 CAR    0          4
  24 ESC    -          -
  25 LIT    -          1
  26 ALM    0          5
  27 CAR    0          5
  28 ESC    -          -
  29 LIT    -          2
  30 ALM    0          3
  31 CAR    0          3
  32 CAR    1          3
  33 OPR    -         10
  34 SAC    -         50
  35 CAR    0          4
  36 CAR    0          5
  37 OPR    -          2
  38 ALM    1          4
  39 CAR    1          4
  40 ESC    -          -
  41 CAR    0          4
  42 ALM    0          5
  43 CAR    1          4
  44 ALM    0          4
  45 CAR    0          3
  46 LIT    -          1
  47 OPR    -          2
  48 ALM    0          3
  49 SAL    -         31
  50 CAR    0          4
  51 CAR    0          5
  52 OPR    -          2
  53 ALM    1          4
  54 RET    -          -
  55 INS    -          5
  56 LEE    -          -
  57 ALM    0          3
  58 LLA    -          1
  59 CAR    0          4
  60 ESC    -          -
  61 RET    -          -
]]>  </ensamblador>
  <fuente>
<![CDATA[(*
    Cálculo de los números de la serie de fibonacci:
    f_0 = 1
    f_1 = 1
    f_n = f_{n-1} + f_{n-2}, n>1
*)
const fib_0=1, fib_1=1;
var n, f;

procedure fibonacci;
    var i,   (* Variable para hacer el recorrido *)
        f_1, (* Número Fibonacci anterior *)
        f_2; (* Número Fibonacci anterior al anterior *)
    begin
        if n=0 then f:=fib_0;
        if n=1 then begin
            f:=fib_1;
            write f; (* Los primeros dos elementos son iguales *)
        end;
        if n>1 then begin
            f_1:=1;
            write f_1;
            
            f_2:=1;
            write f_2;
            
            i:=2;
            while i<n do begin
                f:=f_1+f_2;
                write f;
                f_2:=f_1;
                f_1:=f;
                i:=i+1;
            end;
            f:=f_1+f_2;
        end;
    end;

begin
    read n;
    call fibonacci;
    write f;
end.
]]>  </fuente>
</codigo_pmas>
\end{lstlisting}

\end{document}